\documentclass[doublespace,a4paper,12pt]{article}

\usepackage{amssymb,amsthm,amscd,geometry,fancyhdr,hyperref}
\usepackage{amsmath,amsfonts,setspace,indentfirst,epsfig,slashed}
\usepackage[english]{babel}
\usepackage{graphicx}
\usepackage{mathtext}
\usepackage{indentfirst}
\usepackage{epsfig,amsmath,amsfonts}
\usepackage{mathtools}
\usepackage{color}
\usepackage[usenames,dvipsnames,svgnames,table]{xcolor}
 \usepackage[ normalem]{ulem}

\numberwithin{equation}{section}

\def\a{\alpha}
\def\b{\beta}
\def\g{\gamma} 
\def\d{\delta} \def\e{\epsilon}
  \def\h{\eta} \def\q{\theta}
    \def\m{\mu}
\def\n{\nu}  
 \def\r{\rho}
 \def\s{\sigma}

   \def\Q{\Theta}
   \def\L{\Lambda} 
    
 \def\S{\Sigma}

\def\fr{\frac}  \def\dt{\partial}

\def\ph{\phantom}
\def\mc{\mathcal}

\def\bX{{\mathbb X}}

\newcommand{\br}[2]{\bar{#1}\bar{#2}}
\newcommand{\vW}[4]{W^{\bar{#1}\bar{#2}}_{#3 #4}}
\newcommand{\uW}[4]{W^{#1 #2}_{\bar{#3}\bar{#4}}}
\newcommand{\dlt}[4]{\d^{\bar{#1}\bar{#2}}_{\bar{#3}\bar{#4}}}

\newcommand{\mb}[1]{\mathbf{#1}}

\def\beq{\begin{equation}}
\def\eeq{\end{equation}}
\def\bea{\begin{eqnarray}}
\def\eea{\end{eqnarray}}
\def\ba{\begin{align}}
\def\ea{\end{align}}
\def\pl{\partial}
\def\non{\nonumber}
\def\ab{{\bar{a}}}
\def\bb{{\bar{b}}}
\def\cb{{\bar{c}}}
\def\db{{\bar{d}}}
\def\eb{{\bar{e}}}
\def\fb{{\bar{f}}}
\def\gb{{\bar{g}}}
\def\hb{{\bar{h}}}

\begin{document}

\begin{titlepage}

\vfill
\begin{flushright}
QMUL-PH-12-12
\end{flushright}

\vfill

\begin{center}
   \baselineskip=16pt
   {\Large \bf Duality Invariant M--theory:  \\ Gauged supergravities and Scherk--Schwarz reductions}
   \vskip 2cm
     David S. Berman$^\star$\footnote{\tt d.s.berman@qmul.ac.uk}, Edvard T. Musaev$^\star$\footnote{\tt e.musaev@qmul.ac.uk} and Daniel C. Thompson$^\dagger$\footnote{\tt dthompson@tena4.vub.ac.be}
       \vskip .6cm
             \begin{small}
                          {\it $^\star$Queen Mary University of London, Centre for Research in String Theory, \\
             School of Physics, Mile End Road, London, E1 4NS, England} \\ 
\vspace{2mm}
{\it $^\dagger$ Theoretische Natuurkunde, Vrije Universiteit Brussel and\\ The International Solvay Institutes \\
Pleinlaan 2, B-1050, Brussels, Belgium} \\            
\end{small}
\end{center}

\vfill 
\begin{center} 
\textbf{Abstract}
\end{center} 
\begin{quote}

We consider the reduction of the duality invariant approach to M--theory by a U--duality group valued Scherk--Schwarz twist.  The result is to produce potentials for gauged supergravities that are normally associated with non--geometric compactifications.  The local symmetry reduces to gauge transformations with the gaugings exactly matching those of the embedding tensor approach to gauged supergravity.  Importantly, this approach now includes a nontrivial dependence of the fields on the extra coordinates of the extended space.

\end{quote} 
\vfill
\setcounter{footnote}{0}
\end{titlepage}
\tableofcontents

\section{Introduction} 

Recently, the relationship between the Scherk--Schwarz \cite{Scherk:1979zr} reduction of $O(d,d)$ invariant double field theory (DFT) {\cite{Hull:2009mi,Hohm:2010jy,Hohm:2010pp,Hohm:2010xe,Hohm:2011si,Hohm:2011nu,Jeon:2010rw,Jeon:2011cn,Jeon:2011sq} and gauged supergravities has been  explored in \cite{Aldazabal:2011nj,Geissbuhler:2011mx,Grana:2012rr}.   One central motivation for the double field theory approach to string theory is to make {\it{non--geometric}} backgrounds in some sense geometric, essentially using generalised geometry or its extension.   Whilst the idea of dimensionally reducing on so called non--geometric backgrounds has been extensively explored for years, see for instance \cite{Flournoy:2004vn,Shelton:2005cf,Hull:2005hk,Dabholkar:2005ve,Hull:2006tp,Wecht:2007wu,Grana:2008yw,Hull:2009sg}  and more recently \cite{Dibitetto:2012ia,Dibitetto:2012rk}  it becomes very natural within the context of DFT.  Indeed, by performing a Scherk--Schwarz reduction of DFT \cite{Aldazabal:2011nj,Geissbuhler:2011mx,Grana:2012rr}  succeed in giving a higher dimensional origin to gauged supergravities in which  non-geometric fluxes become purely geometric from the point of view of the doubled space.  

A key feature of double field theory is that there is a constraint (known as the {\it{strong constraint}} or  {\it{physical section condition}})  that reduces the dynamics from the doubled space to a subspace whose dimensionality matches that of physical space time. In general, the consistency of the theory requires all the dynamical fields of the theory  obey the  strong constraint\footnote{Intriguingly, massive IIA supergravity can be accommodated in a consistent way when the section condition is relaxed on the RR sector  \cite{Hohm:2011cp}.}.  This, in turn means that though the double field theory is a revealing rewriting of ordinary supergravity in which the $O(d,d)$ symmetry is manifest, it is, at least locally, completely equivalent to ordinary supergravity. 
 
However,  in \cite{Aldazabal:2011nj,Geissbuhler:2011mx,Grana:2012rr} it is shown that for backgrounds satisfying an appropriate Scherk--Schwarz ansatz the strong constraint is sufficient but not necessary for the reduced theory to be consistent; the constraint can be consistently weakened on the internal space.  Thus, there are consistent DFT backgrounds that are not contained in ordinary supergravity.  The construction of such backgrounds is interesting in its own right since it would explore the structure of DFT beyond supergravity.   
 
All these ideas have extensions to the full U--duality group and M--theory. The action for the M--theory equivalent of double field theory has been written down in a series of works (depending on the specific U--duality group) \cite{Hillmann:2009ci,Berman:2010is,Berman:2011pe,Berman:2011jh,Berman:2011cg}. This builds on earlier work which described the metric of the extended M--theoretic geometry \cite{Hull:2007zu,  Pacheco:2008ps} with recent applications  described in \cite{Malek:2012pw}. The study of the non--geometric backgrounds in M--theory with U--duality twists is given in
\cite{Hull:2006tp} and ideas on the Matrix theory origin of non-geometric backgrounds in M-theory is given in \cite{Chatzistavrakidis:2012qj}.

In this paper we will consider the Scherk--Schwarz reduction along similar line to \cite{Aldazabal:2011nj,Geissbuhler:2011mx,Grana:2012rr}  but applied to the M--theory extended geometry. The M--theory version is a little more involved since the U--duality group jumps with dimension. In dimension $d$, the U--duality group is the exceptional group $E_d$ where for $d<6$ we interpret $E_5=SO(5,5)$ and  $E_4=SL(5)$ etc. For $d>8$ one encounters the infinite-dimensional Kac--Moody type algebras $E_9,E_{10},E_{11}$. In order to be concrete and explicit we will work mostly with the $SL(5)$ U--duality group and its corresponding generalised geometry.  In the final section we will discuss the other U--duality groups.

Since the formalism hasn't been fully developed for $E_{11}$ we cannot, at present, treat the full eleven dimensional theory in a duality symmetric manner. Instead we make the split into a four-dimensional space (which would be the the internal manifold of a traditional compactification) and a seven-dimensional external space-time.  We will neglect the external space-time and work entirely with the four dimensions where the U--duality acts.  In the duality invariant approach we replace this four dimensional internal space with its  associated generalised geometry which is ten dimensional (the coordinates transform in a {\bf 10} of $SL(5)$).  We then cary out a Scherk--Schwarz reduction on this ten dimensional generalised space associated to the SL(5) U--duality group.  We will show that this produces the potential in the remaining seven dimensional space that will be of the form associated with a gauged supergravity. 

A key point will be to check the consistency of the Scherk--Schwarz ansatz with the section condition and the Courant algebra of local symmetries.   We will see that the local symmetries generated by a Dorfman or generalised Lie derivative reduces to gauge transformations. The corresponding gaugings can be associated with the embedding tensor of d=7 gauged supergravity  \cite{Samtleben:2005bp, LeDiffon:2008sh}.  Closure of the gauge algebra gives rise to the known quadratic constraint on the gaugings.  We find that whilst the strong constraint is sufficient for closure it is not necessary. Instead it is replaced with constraints on the generalised geometric fluxes so that they take on a group structure.\footnote{Actually, as we will see this is an abuse of terminology; the ``structure constants'' contain some symmetric pieces as is well known in gauged supergravities.}

The idea that gauged supergravities have higher dimensional origins beyond that of usual supergravity is certainly not new; the work of Riccioni and West \cite{Riccioni:2007ni} considers the application of $E_{11}$ techniques to gauged supergravity and  the tensor hierarchy considerations of de Wit, Nicolai and Samtleben \cite{deWit:2008ta} led to their proposal that that gauged supergarvities probe M-theoretic degrees of freedom beyond usual supergravity.   In this work we reinforce these ideas in the context of the duality invariant M-theory action and the geometry of the extended space.

The the paper is organised as follows. In section 2, we review the generalised geometry for M--theory, its dynamics and its local symmetries.  In section 3, we consider the Scherk--Schwarz reduction at the level of the algebra and its relation to the embedding tensor.  In section 4, we perform the reduction of the action and identify the resulting constraints and potential. Finally, in section 5 we move on to consider briefly the other U--duality groups in particular $SO(5,5)$ and $E_6$.   We include a short appendix with some details on $SL(5)$ and the embedding tensor with trombone gauging. 
\\
\\
{\bf A note on notation:} We will use capital Roman indices, $\mathbb{X}^M,$ to denote the coordinate representation of the generalised space time (i.e. the {\bf 10} of $SL(5)$),  lower case Roman indices $X^m$ to denote the fundamental representation of the duality group (i.e. the {\bf 5} of $SL(5)$) and Greek indices $x^\mu$ denote the coordinates of physical space time. Barred indices denote flattened/tangent/group indices in the corresponding representation.   
\section{Generalised geometry for M--theory}

What follows is a brief summary of the salient details of constructing the generalised geometry for the $SL(5)$ U--duality group in M--theory, see the early seminal work of  \cite{Duff:1990hn} and the developments by \cite{Hull:2007zu,Pacheco:2008ps} and then \cite{Berman:2010is} for the construction of this space and metric.

The duality group acts along 4 directions, we use the indices $\m,\n=1..4$ to denote these directions. The only bosonic supergravity fields active in the four-dimensional space are the metric $g_{\m\n}$ and 3--form $C_{\m\n\r}$ (the six--form potential obviously cannot lie entirely within the four-dimensional space). We truncate all field in the off diagonal space i.e. those with indices in both the four dimensional space and the seven dimensional space are set to zero as are all fermions.

The generalised space is constructed by considering the usual space augmented by the space of membrane windings. The dimension of the space of membrane windings is just the number of two cycles on a four torus which is six. Thus the total space will be ten dimensional, described by 4 usual coordinates, $x^\m$ and 6  winding coordinates $y_{\m\n}=y_{[\m \n]}$.\footnote{In higher dimensions there would also be fivebrane windings, but these only come in for dimension of five and above as discussed in the final section}

Thus the generalised geometry in this case is given by the following tangent space:
\begin{equation}
\label{tan}
TM \oplus \L^2 T^{\ast} M
\end{equation}
that is direct sum of the ordinary tangent space and the space of 2--covectors. Diffeomorphisms and gauge transformations are naturally embedded in this picture as reparametrisations of coordinates on the tangent space \eqref{tan}, this will be important later.

In the extended geometry of M--theory and in double field theory, the tangent space is extended simply as a consequence of extending the space itself. Thus new coordinates are introduced describing the additional dimensions and the generalised geometry of the extended tangent space is just the tangent space of the extended space. (To make the correspondence exact one must impose the solution to the {\it{section condition}} where there is no dependence of any of the fields on the new coordinates. This then relates the generalised geometry construction to the extended space construction; i.e. generalised geometry is extended geometry subject to the trivial solution to the section condition).

We can then write the new extended generalised coordinates on this space in an $SL(5)$ covariant form:
\begin{equation}
\bX^{M }= \bX^{ab}=\left\{
\begin{array}{l}
\bX^{\m 5}=x^\m\\
\bX^{5\m}=-x^\m\\
\bX^{\m\n}=\fr12\h^{\m\n\a\b}\m_{\a\b}
\end{array}
\right.,
\end{equation}
where $\h_{\m\n\a\b}$ is totally antisymmetric symbol $\h_{1234}=1$, $\h^{1234}=1$.  These coordinates lie in a ${\bf{10}}$ of $SL(5)$.

The physical section condition is a quadratic differential condition acting on all fields and their products. Its solution takes you from the extended space to the physical subspace. Its derivation for the $SL(5)$ case was given in \cite{Berman:2011cg}. It lies in a ${\bar{\bf{5}}}$ of $SL(5)$:
 \beq
 \label{WSCn=4}
\frac{1}{4} \e^{a M N} \partial_{M} A \partial_{N} B  \equiv \e^{a b c d e} \partial_{b c} A \partial_{d e} B = 0 \  .
 \eeq
(Note, the somewhat bizarre notation where we have defined the epsilon tensor with mixed indices, this is just another way of writing the usual 5-d epsilon tensor.) For later use we note that the constraint implies anti-symmetrised products of derivatives vanish, 
\begin{equation}
\dt_{[ef}\dt_{cd]}\bullet=0,\quad\dt_{[ef}\bullet\dt_{cd]}\bullet=0 \label{sc2} \ , 
\end{equation}
where $\bullet$ denotes any $SL(5)$ covariant expression.

The generalised diffeomorphism group which combines both usual diffeomorphisms and gauge transformations is generated by a generalised Lie derivative \cite{Berman:2011cg}:
\begin{equation}
\mc{L}_X V^{ab}=\fr12X^{cd}\dt_{cd}V^{ab}+\fr12V^{ab}\dt_{cd}X^{cd}+V^{ac}\dt_{cd}X^{db}-V^{bc}\dt_{cd}X^{da} \ .
\end{equation}
Generalised Lie derivative is an apt name since the derivative can be expressed as a group theoretic modification to the standard Lie derivative
\beq
\label{sl5deriv}
{\cal L}_X V^M = X^N\pl_N V^M - V^N\pl_N X^M + \e_{a PQ} \e^{a MN} \pl_N X^P V^Q \ .
\eeq

The anti-symmetrisation of these generalised Lie derivative gives rise to a so called generalised C--bracket, 
\begin{equation}
 \label{gcb}
[X,Y]_C^{ab}=\fr12(\mc{L}_XY-\mc{L}_YX)^{ab}=\fr14X^{cd}\dt_{cd}Y^{ab}+\fr14Y^{ab}\dt_{cd}X^{cd}+\fr12Y^{c[a}\dt_{cd}X^{b]d}-(Y\leftrightarrow X) \ ,
\end{equation}
whose projection under $SL(4)$ is the Courant bracket. 

The algebra of generalised transformations closes on the generalised {\it{C--bracket}} up to the strong constraint, 
\begin{equation}
[\d_{X_1},\d_{X_2}]Q^{ab}=[\mc{L}_{X_1},\mc{L}_{X_2}]Q^{ab}=\mc{L}_{[X_1,X_2]_C}Q^{ab}+F_0^{ab} \ .
\end{equation}
The extra term $F_0$ that violates the closure has the form \cite{Berman:2011cg}
\begin{equation}
\begin{split}
\label{F_0}
F_0^{ab}=&\fr34\left(X_1^{ef}\dt_{[ef}X_2^{cd}\dt_{cd]}Q^{ab}+Q^{ab}X_1^{ef}\dt_{[ef}\dt_{cd]}X_2^{cd}+ Q^{ab}\dt_{[ef}X_1^{ef}\dt_{cd]}X_2^{cd}-\right.\\
&\left.-4Q^{c[a}\dt_{[ef}X_1^{b]d}\dt_{cd]}X_1^{ef}+2X_1^{ef}Q^{c[a}\dt_{[ef}\dt_{cd]}X_2^{b]d}+ 2Q^{c[a}X_1^{b]d}\dt_{[ef}\dt_{cd]}X_2^{ef}\right) - X_1 \leftrightarrow X_2 \ ,
\end{split}
\end{equation}
and since all derivatives appear completely antisymmetrised this vanishes once the strong constraint \eqref{sc2} is imposed.

The dynamical fields of supergravity are encoded in a metric on the generalised tangent space introduced in \cite{Hull:2007zu,Pacheco:2008ps,Berman:2010is} of the form\footnote{As explained in \cite{Berman:2011jh} there is freedom to rescale this generalised metric by a power of $\det g$. With the given normalisation the  derivative \eqref{sl5deriv} generates a diffeomorphism on $g_{\m \nu}$. With a different normalisation one would need to add some appropriate density term to  \eqref{sl5deriv} and also modify the coefficients appearing in the potential \eqref{lag}.} 
\begin{equation}
\label{genmet}
M_{M  N} = M_{ab,cd}=
\begin{bmatrix}
M_{\m 5,\n 5} & M_{\m 5,\a\b} \\
M_{\g\d,\n 5} & M_{\g\d,\a\b}
\end{bmatrix}=
\begin{bmatrix}
g_{\m\n}+\fr12 C_\m^{\ph{\m}\r\s}C_{\n\r\s} & -\fr{1}{2\sqrt{2}}C_\m^{\ph{m}\r\s}\h_{\r\s\a\b}\\
-\fr{1}{2\sqrt{2}}C_\n^{\ph{m}\r\s}\h_{\r\s\g\d} & g^{-1}g_{\g\d,\a\b} 
\end{bmatrix} \ , 
\end{equation}
in which $g=\det{g}$ and $g_{\a\b,\m\n}=\fr12(g_{\a\m}g_{\b\n}-g_{\a\n}g_{\b\m})$. Since this metric can be parametrised by  $g_{\m\n}$ and  the 3--form $C_{\m\n\r}$ it contains only $10+4=14$ independent components. This is in accordance with the fact that generalised metric $M_{ab,cd}$ is a representative of the coset space $SL(5)/SO(5)$ whose dimension is  $24-10=14$.

The dynamics of the generalised space are given by a potential:
\beq
\label{lag}
V = M^\alpha  \left(  c_1 V_1 + c_2 V_2 + c_3 V_3 +c_4 V_4  \right)
\eeq
where
\begin{equation}
\begin{split}
& V_{1} = M^{MN} \pl_M M^{KL} \pl_N M_{KL} \ ,  V_2 =  M^{MN} \pl_M M^{KL} \pl_K M_{NL}  \ ,  \\ 
& V_{3} = - \pl_M M^{MQ}\left( M^{RS} \pl_P M_{RS} \right)  \ ,  V_{4} = M^{MN}  \left( M^{RS} \pl_M M_{RS} \right) \left( M^{KL} \pl_N M_{KL} \right) \ ,
\end{split}
\end{equation}
and with the generalised metric given by \eqref{genmet} the constants are 
\beq
 \a = -1/4 \ , \quad  c_1 = \frac{1}{12} \ , \quad c_{2} = - \frac{1}{2} \ , \quad  c_3 = \frac{1}{4} \ , \quad c_4 = \frac{1}{12} \ . 
\eeq
Note, the power of det(M) in the above is simply given by noting that, ${\rm{det}}(g)={\rm{det}}(M)^{-2}$. 
Of course, in the full theory one should supplement this with a Lagrangian for the seven dimensional external space though we do not do so here.

\section{Scherk--Schwarz formalism}

For the case at hand we have the $SL(5)$ U--duality group for a 4--dimensional compact manifold. We have split the eleven dimensional space into a direct product $M_4\times M_7$ with $M_4$ a compact 4--dimensional manifold and $M_7$ some 7--dimensional manifold that will not play a role. In the framework of generalised geometry for \mbox{M--theory} one focuses only on compact dimensions $M_4$ where the U--duality group acts and replaces it with the associated generalised geometry.

We will now carry out a Scherk--Schwarz reduction on the generalised geometry to give a potential in the remaining seven dimensions.  We separate out the dependence on internal coordinates by introducing twisting matrices \cite{Scherk:1979zr}:
\begin{equation}
\label{W}
Q^{ab}(x_{(7)},\bX)=W^{ab}_{\br{a}{b}}(\bX)Q^{\br{a}{b}}(x_{(7)}), \quad 
\end{equation}
where $x_{(7)}$ denotes the dependance on the coordinates on $M_7$ (which we henceforth suppress).    These twisting matrices are valued in are in a ${\bf{10 \times 10}}$ representation but can be decomposed in terms of ${\bf{5 \times 5}}$
\begin{equation}
\label{WtoV}
W^{ab}_{\br{a}{b}}=\fr12\left[V^a_{\bar{a}}V^b_{\bar{b}}-V^a_{\bar{b}}V^b_{\bar{a}}\right] \ . 
\end{equation}
With the U--duality group $SL(5)$ it is natural to impose that $\det V = 1 $, however the generalised geometry for M--theory actually also supports a natural $\mathbb{R}^+$ action  \cite{Coimbra:2011ky} so for the time being we shall not do so.

\subsection{Twisted derivatives and gaugings}

In the usual Scherk--Schwarz reduction, the structure constants of the gauge group in the reduced theory are obtained by the twisting of Lie derivative in the parent theory. Thus we will now examine the twisted generalised Lie derivative obtained by invoking the Scherk--Schwarz ansatz on all fields and gauge parameters.  The result is 
\begin{equation}
\label{Lie}
(\mc{L}_\S Q)^{ab}=W^{ab}_{\br{a}{b}}(\bX)\left[(\mc{L}_{\bar{\S}}\bar{Q})^{\br{a}{b}} +X_{\br{c}{d},
\br{e}{f}}{}^{\br{a}{b}}\S^{\br{c}{d}}Q^{\br{e}{f}}\right] =: W^{ab}_{\br{a}{b}}\left(\bar{\mc{L}}_{\bar{\S}}Q\right)^{\br{a}{b}},
\end{equation}
with would-be structure constants given in terms of the twist matrices by
\bea
\label{F}
X_{\br{c}{d},\br{e}{f}}{}^{\br{a}{b}}&=& \frac{1}{2} \left(   \vW{a}{b}{m}{n} \partial_{\bar{c} \bar{d}} \uW{m}{n}{e}{f} -   \vW{a}{b}{m}{n} \partial_{\bar{e} \bar{f}} \uW{m}{n}{c}{d}   +\frac{1}{4} \epsilon^{ \bar{a} \bar{b} \bar{i} \bar{j} \bar{k}}\epsilon_{\bar{k} \bar{p} \bar{q} \bar{c} \bar{d} }  \vW{p}{q}{m}{n} \partial_{\bar{i} \bar{j}} \uW{m}{n}{a}{b}  \right) \nonumber  \\ 
&=& \fr12\vW{a}{b}{m}{n}\uW{p}{q}{c}{d}\dt_{p q}\uW{m}{n}{e}{f}+\fr12\dlt{a}{b}{e}{f}\dt_{mn}\uW{m}{n}{c}{d} +2\vW{a}{b}{m}{n}\uW{m}{p}{e}{f}\dt_{pq}\uW{q}{n}{c}{d} \ . 
\eea
We encounter our first constraint on the Scherk--Schwarz twist element which is that these objects are constant. However, an immediate difference to the $O(d,d)$ case is that these ``structure constants" are not anti-symmetric in their lower indices -- to correct this misnomer we shall refer to them as gaugings rather than structure constants.  By making use of the invariance of the epsilon tensor and the decomposition   \eqref{WtoV}, the symmetric part of the gaugings can be extracted as
\beq
X_{\br{c}{d},\br{e}{f}}{}^{\br{a}{b}}+X_{\br{e}{f},\br{c}{d}}{}^{\br{a}{b}} = \frac{1}{8} \epsilon_{\bar{i} \bar{c} \bar{d} \bar{e} \bar{f} } \epsilon^{\bar{j} \bar{m} \bar{n} \bar{a}\bar{b}} V^{\bar{i}}_{p} \partial_{\bar{m} \bar{n}} V^p_{\bar{j}} \  .
\eeq
To see the full content of the gauging it is in fact  helpful to decompose according  \eqref{WtoV}.  One finds that 
\beq
X_{\br{c}{d},\br{e}{f}}{}^{\br{a}{b}}  = 2 X_{\bar{c}\bar{d} , [\bar{e}}{}^{[\bar{a}}\delta_{\bar{f}]}^{\bar{b}]}  \ ,
\eeq
with
\beq
X_{\bar{c}\bar{d} , \bar{e}}{}^{\bar{a}} = \frac{1}{2} V^{\bar{a}}_{m} \partial_{\bar{c} \bar{d} } V_{\bar{e}}^m + (T^{\bar{p}}_{\bar{q}})^{\bar{m} \bar{n}}_{\bar{r} [\bar{c}} \left( V^{\bar{r}}_t \partial_{\bar{m} \bar{n}} V^t_{\bar{d}]}\right) (T^{\bar{q}}_{\bar{p}})^{\bar{a}}_{\bar{e}} - \frac{1}{10} \delta^{\bar{a}}_{\bar{e}} V^{\bar{m}}_t \partial_{\bar{m} [\bar{c}} V_{\bar{d}]}^t \ ,
\eeq
in which  $(T^{\bar{q}}_{\bar{p}})^{\bar{a}}_{\bar{e}}$ and $(T^{\bar{p}}_{\bar{q}})^{\bar{m} \bar{n}}_{\bar{r} \bar{c}} $ are the $SL(5)$ generators in the $\bf{5}$ and $\bf{10}$ respectively (see appendix).  This result can be expressed as 
\beq
X_{\bar{c}\bar{d} , \bar{e}}{}^{\bar{a}} = \delta^{\bar{a} }_{[\bar{c} } Y_{\bar{d}] \bar{e} }  - \frac{10}{3} \delta^{\bar{a} }_{[\bar{c} }  \theta_{\bar{d}] \bar{e} }   - 2 \epsilon_{\bar{c}\bar{d} \bar{e} \bar{m} \bar{n} } Z^{\bar{m} \bar{n}, \bar{a} }  + \frac{1}{3} \theta_{\bar{c}\bar{d}} \delta_{\bar{e}}^{\bar{a} } \ ,
\eeq
where $Y_{\bar{c} \bar{d}}= Y_{\bar{d} \bar{c}}$ is in the $\bf{15}$ and is given by 
\beq
Y_{\bar{c} \bar{d} } =  V^{\bar{m}}_t \partial_{\bar{m} (\bar{c} } V^t_{\bar{d})}  \ , 
\eeq
and $ Z^{\bar{m} \bar{n}, \bar{p} } = -  Z^{\bar{n} \bar{m}, \bar{p} } $ is in the $\overline{\bf{40}}$ such that  $ Z^{[\bar{m} \bar{n}, \bar{p}] } = 0 $  is given by 
\beq
Z^{\bar{m} \bar{n}, \bar{p} } = - \frac{1}{24} \left( \epsilon^{\bar{m} \bar{n} \bar{i} \bar{j} \bar{k} } V_{t}^{\bar{p}} \partial_{\bar{i}\bar{j}} V_{\bar{k}}^t    + V_{t}^{[\bar{m}} \partial_{\bar{i}\bar{j}} V^{|t|}_{\bar{k}}  \epsilon^{\bar{n}]  \bar{i} \bar{j} \bar{k}  \bar{p}}  \right)  \ ,
\eeq
and $\theta_{\bar{c} \bar{d}}= - \theta_{\bar{d} \bar{c}}$ is in the $\bf{10}$ and is given by 
\beq
\theta_{\bar{c} \bar{d}}  =  \frac{1}{10 } \left( V^{\bar{m}}_t \partial_{\bar{c} \bar{d}} V_{\bar{m}}^t  - V^{\bar{m}}_t \partial_{\bar{m} [\bar{c} } V^t_{\bar{d}]} \right)  \ . 
\eeq
It is note worthy that although $\bf{10 \otimes 24}=  \bf{10 \oplus 15 \oplus \overline{40} \oplus  175}$ the $\bf{175}$ makes no appearance in the gaugings produced by Scherk--Schwarz reduction.  The reason for this will become apparent momentarily but first let us consider the constraints that come from closure of the algebra.   

\subsection{Closure and quadratic constraints}
We now study the conditions placed on the Scherk--Schwarz twist by demanding closure of the algebra.  In this section we will find it expedient to use indices in the $\bf{10}$.   We recall the un-twisted closure is given by 
\begin{equation}
\mc{L}_{[X_1,X_2]_C} Q^M -  [\mc{L}_{X_1},\mc{L}_{X_2}]Q^{M } = -  F_0^{M} \ ,
\end{equation}
where the anomalous term $F_0^M$ vanishes on the section condition.  One way to proceed would be to explicitly evaluate $F_0^M$ with the gauge parameters obeying Scherk--Schwarz ansatz.   Instead we simply evaluate the left hand side making use of the result that 
\beq
\label{lred}
{\cal L_{\xi}} Q^M = W^M_{\bar{C}} X_{\bar{A} \bar{B} }{}^{\bar{C} } \xi^{\bar{A}} Q^{\bar{B}} 
\eeq
and noting that $X_{\bar{A} \bar{B} }{}^{\bar{C} } ,    \xi^{\bar{A}} , Q^{\bar{B}}$ are all assumed to be constant with respect to the derivative $\pl_M$. 
Then one finds the condition for closure becomes 
\beq
\frac{1}{2} \left( X_{\bar{A} \bar{B} }{}^{\bar{C}} -    X_{\bar{B} \bar{A} }{}^{\bar{C} }\right)X_{\bar{C} \bar{E}}{}^{\bar{G}} - X_{\bar{B} \bar{E}}{}^{\bar{C}}X_{\bar{A} \bar{C}}{}^{\bar{G}} +   X_{\bar{A} \bar{E}}{}^{\bar{C}}X_{\bar{B} \bar{C}}{}^{\bar{G}}  = 0 \  . 
\eeq
 If we define $(X_{\bar{A}})_{\bar{B}}^{\bar{C}} = X_{\bar{A} \bar{B} }{}^{\bar{C}}$ this may be written in the suggestive form 
  \beq
  \label{closure1}
 [ X_{\bar{A}}, X_{\bar{B}}]  = - X_{[\bar{A} \bar{B}]}{}^{\bar{C}} X_{\bar{C}}  \ . 
 \eeq
 Thus we begin to see the structure of an algebra of gauge transformation
 \beq
 \delta_{\xi} V^{\bar{C}} = \xi_1^{\bar{A}} (X_{\bar{A}})_{\bar{B}}{}^{\bar{C}} V^{\bar{B}}  \ .
 \eeq 
 
A second closure constraint comes from considering the Jacobiator of gauge transformations.  Making use of the first closure constraint  \eqref{closure1}  we find that 
\beq
\label{Jac}
[\delta_{\xi_1}, [\delta_{\xi_2}, \delta_{\xi_3} ] ] + c.p. = \left( X_{[\bar{A}\bar{B}]}{}^{\bar{E}}X_{[\bar{E} \bar{C}] }{}^{\bar{G}}   +   X_{[\bar{C}\bar{A}]}{}^{\bar{E}}X_{[\bar{E} \bar{B}] }{}^{\bar{G}} +X_{[\bar{B}\bar{C}]}{}^{\bar{E}}X_{[\bar{E} \bar{A}] }{}^{\bar{G}}   \right)X_{\bar{G} \bar{D}}{}^{\bar{F} } \xi_1^{\bar{A}}\xi_2^{\bar{B}} \xi_3^{\bar{C}} V^{\bar{D}} 
\eeq
where $c.p.$ denotes cyclic permutations.  The right hand side of this can be understood as the Jacobi identity for the antisymmetric part of $X$ {\it{projected}} into the algebra generator.   The right hand side needs to vanish for a consistent algebra but we emphasise that the Jacobi identity for $X_{[AB]}{}^C$ needs only to hold after projection. 

A final closure constraint comes from understanding that the $X_{MN}{}^K$ should be not only constant but also invariant objects under the local symmetry transformations.
We will see shortly that this is also necessary  so that the reduced action does not depend on the internal coordinates and for it to be gauge invariant.  Since
\beq
\delta_{\xi} X_{\bar{A} \bar{B}}{}^{\bar{C}}  =   \xi^{\bar{E}} \left(  [ X_{\bar{E}}, X_{\bar{A}}] _{\bar{B}}{}^{\bar{C}} +  X_{\bar{E} \bar{A}}{}^{\bar{D}} (X_{\bar{D}})_{\bar{B}}{}^{\bar{C}} \right) 
\eeq
we conclude that in addition to the first closure constraint \eqref{closure1}  that the symmetric part $X_{(\bar{A}\bar{B})}{}^{\bar{C}}$ must vanish when projected into a generator 
\beq
X_{(\bar{A} \bar{B})}{}^{\bar{C}} X_{\bar{C}}  = 0 \ ,
\eeq
so that 
 \beq
  \label{closure2}
 [ X_{\bar{A}}, X_{\bar{B}}]  = - X_{\bar{A} \bar{B}}{}^{\bar{C}} X_{\bar{C}}  \ . 
 \eeq
 In fact, this final constraint is enough to guarantee that the Jacobiator of gauge transformations   \eqref{Jac}  vanishes.   This can be seen by considering the Jacobi identity for the commutator appearing in  \eqref{closure2}.

 \subsection{The embedding tensor and gauged supergravity}

Toriodal compactification of eleven-dimensional supergravity gives rise to maximal supergravitives in $D=11- d$ dimensions which admit a global $G=E_{d(d)}$ duality (or Cremmer-Julia) symmetry.  These theories allow susy preserving deformations, known as gaugings, in which some subgroup of the global $E_{d(d)}$ symmetry is promoted to a local symmetry.  The resultant gauged supergravities have non-abelian gauge groups and develop a potential for the scalar fields.    A  universal approach to gauged supergravities is the embedding tensor (for a review see \cite{Samtleben:2008pe}) which describes how the gauge group generators are embedded into the global symmetry.  Treated as a supurionic object the embedding tensor provides a manifestly $G$ covariant description of the gauged supergravities. 

  In addition to the global $E_{d(d)}$ symmetry the toriodially reduced theories also posses a global $\mathbb{R}^+$ scaling symmetry  know as the trombone symmetry (this is an on-shell symmetry for $D\neq2$).   This gives rise to a more general class of gaugings whereby a subgroup of the full $G\times \mathbb{R}^+$  is promoted to a local symmetry.  The embedding tensor approach was extended to incorporate such trombone gaugings in \cite{LeDiffon:2008sh}.
  
  We specialise to the case of $D=7$ \cite{Samtleben:2005bp} for which the vector fields of the un-gauged/abelian theory are in the $\overline{\bf{10}}$ of $SL(5)$ and the two-forms are in the $\bf{5}$ and the scalars parametrise an $SL(5)/SO(5)$ coset.   The gaugings are specified by the embedding tensor $\widehat{\Theta}_{mn , p }{}^q$ that  projects generators  $\{T^a_b\}$ of global group\footnote{Traceless matrices among $\{T^a_b\}$ correspond to $SL(5)$ generators and the trace part gives generator of $\mathbb{R}^+$ trombone symmetry.}  $SL(5)\otimes\mathbb{R}^+$ to some subset $X_{mn}=\widehat{\Q}_{mn, b}{}^a T^b_a$ which generate the gauge group and enter into covariant derivatives:
  \beq
  D = \nabla - g A^{mn} X_{m n} \ .
    \eeq 

 {\it{A priori}} the embedding tensor is in the $\mathbf{10\otimes24}$ representation of $SL(5)$ that decomposes as
\begin{equation}
\widehat{\Theta}_{mn, b}{}^a\in\mathbf{10\otimes24}=\mb{10\oplus15\oplus\overline{40}\oplus175}.
\end{equation}
However, preservation of supersymmetry gives a linear constraint restricting the embedding tensor only to $\mb{10}\oplus\mb{15\oplus\overline{40}}$ which yields (see appendix for full details)
\begin{equation}
\begin{split}
&\widehat{\Q}_{mn, b}{}^a=\Q_{mn , b }{}^a +\q_{mn}\d^a_b,\\
&\Q_{mn, b}{}^a=\d^a_{[m}Y_{n]b}- 2 \e_{mnrs b}Z^{rs,a}- \frac{5}{3} \q_{kl}(T^a_b)^{kl}_{mn},
\end{split}
\end{equation}
where  $\theta, Y$ and $Z$ are in the $\bf{10, 15}$ and $\bf{\overline{40}}$ respectively. The components of the embedding tensor in the $\bf{10}$ are associated to the trombone gaugings.  
     
The gauge group generators in the $\mb{5}$ and $\mb{10}$ representations have the following form
\begin{equation}
(X_{mn})^a_b=\widehat{\Theta}_{mn}{}^a_b \ , \quad
(X_{mn})_{pq}^{rs}=2X_{mn}{}^{[r}_{[p}\d^{s]}_{q]}.
\end{equation}
and it is now evident that the gaugings obtained in the previous section are in complete agreement with the above.  The fact that the $\bf{175}$ was automatically absent in the reduction of the extended geometry means the linear constraint is satisfied automatically, i.e. the embedding tensor appears already projected on corresponding representation.  This can be understood as the compatibility of the generalised Lie derivative with supersymmetry (this property was commented on in 
\cite{Coimbra:2011ky}).  

As was mentioned earlier, in addition to the linear constraint there is also a quadratic constraint which arises by demanding closure of the gauge algebra
\beq
[X_{mn }, X_{pq}] = - ( X_{mn})_{pq}{}^{r s} X_{r s}  \ , 
\eeq
which requires that 
\beq
Z^{m n , p} X_{m n} = 0 \ . 
\eeq
This is in exact agreement with the closure constraints found in the preceding section.

\subsection{Relation to DFT}
To close this section we address an obvious question: can one directly relate the above $SL(5)$ twisting to the $O(d,d)$ T--duality twisting in double field of \cite{Aldazabal:2011nj,Geissbuhler:2011mx,Grana:2012rr}.  The answer is of course yes.  Just as type IIA supergravity can be obtained by a dimensional reduction of M--theory,  generalised M--theory reduces to double field theory as reported in \cite{Thompson:2011uw}.   At the level of the derivative it is quite straightforward to reduce the generalised Lie derivative \eqref{sl5deriv} to the $O(3,3)$ case \cite{Berman:2011cg}. One asserts that $\partial_{4 5} = 0$ and $\partial_{\a 4}=0$ for $\a = 1\dots3$.   If we define  $X^\Lambda = ( X^{\a} , \tilde{X}_\a) = (X^{\a 5}, \frac{1}{2} \e_{\a \b \g} X^{\b \g})$ then one recovers from \eqref{sl5deriv} 
\bea
({\cal L}_X V)^{\a} = X^\Lambda \pl_\Lambda V^\a + ( \pl^\a X_\Lambda - \pl_\Lambda X^\a) V^\Lambda \ , \\
\frac{1}{2} \e_{ \a \b \g } ({\cal L}_X V)^{\b \g}=   X^\Lambda \pl_\Lambda \tilde{V}_\a + ( \pl_\a X_\Lambda - \pl_\Lambda X_\a) V^\Lambda \ .
\eea
These are just the components of the $O(3,3)$ generalised derivative for the DFT.  Since this holds before twisting it necessarily holds after twisting.  Thus, our results truncated to the NS sector fields upon dimensional reduction would correspond to the expressions for gaugings and non-geometric fluxes found in  \cite{Aldazabal:2011nj,Geissbuhler:2011mx,Grana:2012rr} and also  \cite{Andriot:2012wx,Andriot:2011uh,Andriot:2012an}.  However this work automatically includes the RR sector as well and thus offers a  way to extend these results and to study generalizations of U-folds and their direct connection to non-geometric fluxes.

\section{Effective potential}

In this section we apply apply the Scherk--Schwarz ansatz to the generalised metric itself and reduce the action to give the effective potential of the reduced theory.

   \subsection{Scherk--Schwarz reduction}

We recall the action is given by 
\beq
V =  \sqrt{g}    \left(  c_1 V_1 + c_2 V_2 + c_3 V_3 +c_4 V_4  \right)
\eeq
where
\begin{equation}
\begin{split}
& V_{1} = M^{MN} \pl_M M^{KL} \pl_N M_{KL} \ , \quad V_2 =  M^{MN} \pl_M M^{KL} \pl_K M_{NL}  \ ,  \\ 
& V_{3} = -\pl_M M^{MP} \left( M^{RS} \pl_P M_{RS} \right)  \ ,\quad V_{4} = M^{MN}  \left( M^{RS} \pl_M M_{RS} \right) \left( M^{KL} \pl_N M_{KL} \right) 
\end{split}
\end{equation}
and with the generalised metric given by \eqref{genmet} the constants are 
\beq
c_1 = \frac{1}{12} \ , \quad c_{2} = - \frac{1}{2} \ , \quad  c_3 = \frac{1}{4} \ , \quad c_4 = \frac{1}{12} \ . 
\eeq
In addition to the above terms we may, with impunity,  also consider the inclusion of terms that vanish identically when the strong constraint is imposed.   The most natural such term to consider would be 
\beq
V_{5} = \epsilon^{a MN} \epsilon_{a PQ}  E^A_{R} M^{RS} E^B_{S}  \partial_{M} E^{P}_A \partial_N E^Q_B \ , 
\eeq
in which $E^A_M$ is a vielbein for $M_{MN} = E^{A}_M \delta_{AB} E^{B}_N$.  In fact, we shall see that such a term is indispensable to match the reduced theory with the potential for gauged supergravity. 

We now apply the Scherk--Schwarz ansatz   to the terms in the action to find the reduced theory.   We will find it convenient to work not with the $10 \times 10$ matrix {\it{big}}  $M_{MN}$ but instead with the $5 \times 5$ {\it{little}} $m_{mn}$ defined by 
\bea
M_{MN}= M_{mn,pq}=    m_{mp} m_{nq} - m_{mq} m_{pq}  \ ,  \nonumber \\
M^{MN}= M^{mn,pq}=   m^{mp} m^{nq} - m^{mq} m^{pq}  \ .
\eea
The metric in the fundamental is given by 
\beq
m_{mn}=  \left( \begin{array}{cc} g^{-1/2}  g_{\mu \nu}  & V_\nu \\
V_\m & \det{g}^{1/2}( 1+V_\m V_\n g^{\m \n}   )  \end{array}\right)
\eeq
where $V^\m = \frac{1}{6} \e^{\m \n \rho \sigma} C_{\nu \rho \sigma}$ and $ \e^{\m \n \rho \sigma} $ is the alternating {\it tensor}. This object has determinant $\det m_{mn} = \det{g}^{-\frac{1}{2} }$.  In terms of little $m$ the terms in the potential reads\footnote{We found the computer algebra package {\tt{Cadabra}} \cite{Peeters:2007wn,Cadabra} a useful tool for verifying some of the more laborious manipulations in this section} 
\bea
V_{1} &=& \frac{3}{2} m^{pr} m^{qs} \pl_{pq} m^{mn} \pl_{rs} m_{mn}   - \frac{1}{2}  m^{pr} m^{qs} Tr(m^{-1}\pl_{pq} m)  Tr(m^{-1}\pl_{mn} m) \ , \nonumber\\ 
V_{2} &=&   m^{pr} m^{qs} \pl_{pq} m^{kl} \pl_{ks} m_{rl} - \pl_{pq} m^{pk} \pl_{kl} m^{lq} \ , \nonumber\\ 
V_{3} &=&4m^{mq}m^{ij} \pl_{pq}m_{ij} \pl_{mk} m^{kp} ,  \nonumber \\ 
V_{4} &=&    8 m^{pr} m^{qs} Tr(m^{-1}\pl_{pq} m)  Tr(m^{-1}\pl_{mn} m)  \  . 
\eea
 For {\it little}  $m$ the Scherk--Schwarz ansatz is then  
\beq
\label{ansatz}
m_{mn} =  V_m^\ab m_{\ab \bb} V_n^\bb     \ . 
\eeq

Let us introduce some notation:
\beq
\Lambda^{\ab}{}_{\bb \cb \db} = V^{\ab}_m \partial_{\bb \cb} V^m_{\db} \ ,  \quad 
\Lambda^{\ab}{}_{\bb \cb \ab} = \chi_{\bb \cb} \ ,  \quad \Lambda^{\ab}{}_{\ab \bb \cb} = \psi_{\bb \cb} \ . 
\eeq
Assuming that $\partial_{mn } m_{\ab\bb} = 0$ we obtain
\bea
 V_{1}  &=& - m^{\ab \bb}  m^{\cb \db} \left[3\Lambda^{\eb}{}_{\ab\cb \fb} \Lambda^{\fb}{}_{\bb \db \eb}   + 2    \chi_{\ab \cb} \chi_{\bb \db}  +3   m_{\eb \fb} m^{\gb \hb} \Lambda^{\eb}{}_{\ab \cb \gb} \Lambda^{\fb}{}_{\bb\db\hb} \right] \ ,   \nonumber \\
V_{2} &=&-  m^{\ab \bb} m^{\cb \db}\left[ 2\Lambda^{\eb}{}_{\fb \ab \cb} \Lambda^{\fb}{}_{\db \bb \eb} +  \Lambda^{\eb}{}_{\fb \ab \cb} \Lambda^{\fb}{}_{\eb \bb \db}  -   \Lambda^{\eb}{}_{\fb \ab \bb} \Lambda^{\fb}{}_{\eb \cb \db} + 2 \psi_{\eb \cb} \Lambda^{\eb}{}_{\db \ab \bb} -\psi_{\ab\cb} \psi_{\bb \db}    \right]\ ,   \nonumber \\
&& \quad -  m^{\ab \bb} m^{\cb \db}\left[  m_{\eb \fb} m^{\gb \hb} \Lambda^{\eb}{}_{\ab \cb \gb} \Lambda^{\fb}{}_{\hb\db\bb}  \right]\ ,   \nonumber \\
V_{3} &=& - 8 m^{\ab \bb} m^{\cb \db} \left[ \chi_{\eb \ab} \Lambda^{\eb}{}_{\bb \cb \db}   + \chi_{\bb \cb} \psi_{\ab \db} \right]\ , \nonumber \\
V_{4} &=& 32 m^{\ab \bb} m^{\cb \db} \chi_{\ab \cb} \chi_{\bb \db} \ ,
\eea
for the original terms in the action.  For the extra term (which vanishes upon the strong constraint) we find 
\beq
\begin{split}
V_{5} &=  - \e^{\ab \bb \cb \db\eb} \e_{\ab  \fb \gb \hb \bar{i} } \left(  m^{\bar{p}  \bar{i} } m^{\bar{q} \gb}    \right) \Lambda^\fb{}_{\bb \cb \bar{p} } \Lambda^\hb{}_{\db \eb \bar{q} }      \\   
 &=    -4 (\psi_{\ab \bb} m^{\ab \bb})^2   + 4  m^{\ab \bb} m^{\cb \db}\left[  \psi_{\ab \cb} \psi_{\bb \db} +2 \psi_{\eb \cb} \Lambda^{\eb}{}_{\db \ab \bb}  +  \Lambda^{\eb}{}_{\fb \ab \bb}    \Lambda^{\fb}{}_{\eb \cb \db}  -    \Lambda^{\eb}{}_{\fb \ab \db}    \Lambda^{\fb}{}_{\eb \cb \bb}   \right]
  .  \end{split}
\eeq

To proceed we shall simplify matters by assuming
\beq
\det g = 1 \ , \quad   \det m = 1 \ ,  \quad   \det V = 1 \ , 
\eeq
 and further that the trombone gauging vanishes.  Then we have the following identifications:
 \beq
 \chi_{\ab \bb} = 0 \  , \quad \psi_{[\ab \bb]} = 0 \ ,  \quad  ,  Y_{\ab \bb}  =  \psi_{(\ab \bb)}  \ , \quad   Z^{\bb \cb , \ab}    = - \frac{1}{16}  \Lambda^{\ab}{}_{\db \eb \fb}\e^{\db \eb \fb \bb \cb}   \ . \eeq

Using the invariance of the $\epsilon$-tensor we then find the following relations: 
\beq
\begin{split}
 64  Z^{\ab \bb, \cb} Z^{\db \eb \fb} m_{\ab \db} m_{\bb \eb} m_{\cb \fb} =& \Lambda^{\ab}{}_{\bb \cb \db} \Lambda^{\eb}{}_{ \fb \gb \hb} m_{\ab \eb } m^{\bb \fb} m^{\cb \gb} m^{\db \hb} - 2 \Lambda^{\ab}{}_{\bb \cb \db} \Lambda^{\eb}_{\fb \gb \hb} m_{\ab \eb} m^{ \bb \fb} m^{ \cb \hb } m^{ \db \gb}  \\
   64   Z^{\ab \bb, \cb} Z^{\db \eb \fb}  m_{\ab \db} m_{\bb \cb} m_{\eb \fb} =& \Lambda^{\ab}{}_{\bb \cb \db} \Lambda^{\bb}{}_{\ab \eb \fb} m^{ \cb \fb} m^{ \db \eb } - \frac{1}{2}\Lambda^{\ab}{}_{\bb \cb \db} \Lambda^{\db}{}_{\eb \fb \ab} m^{\bb \eb} m^{ \cb \fb}  \\
   &   - \Lambda^{\ab}{}_{\bb \cb \db} \Lambda^{\bb}_{\ab \eb \fb} m^{\cb \eb} m^{ \db \fb} - 2 \Lambda^{\ab}{}_{\bb \cb \db} \Lambda^{\bb}{}_{\eb \fb \ab } m^{ \cb \eb } m^{ \db \fb }   \\ 
   & +  \frac{1}{2} \Lambda^{\ab}{}_{\bb \cb \db} \Lambda^{\eb}{}_{ \fb \gb \hb} m_{\ab \eb } m^{\bb \fb} m^{\cb \gb} m^{\db \hb} -  \Lambda^{\ab}{}_{\bb \cb \db} \Lambda^{\eb}_{\fb \gb \hb} m_{\ab \eb} m^{ \bb \fb} m^{ \cb \hb } m^{ \db \gb}       \ .
\end{split}
\eeq

Putting things together we then find that 
\beq
\frac{1}{12} V_1 - \frac{1}{2} V_2 - \frac{1}{8} V_5  =- 32 V_{gauged}  + \Lambda^{\ab}{}_{\bb \cb \db} \Lambda^{\bb}{}_{\ab \eb \fb} \left( m^{\cb \eb } m^{\db \fb}- m^{\cb \db} m^{\eb \fb} \right) 
\eeq
 where $V_{gauged}$ is the known potential for the scalars in gauged supergravity given by \cite{Samtleben:2005bp}:
 \beq
 \label{Vgauged}
V_{gauged} = \frac{1}{64} \left( 2{m}^{\ab \bb}  Y_{\bb \cb} {m}^{\cb \db} Y_{\db \ab} -   ( {m}^{\ab \bb} Y_{\ab \bb})^2   \right) + Z^{\ab \bb, \cb} Z^{\db \eb, \fb} \left( {m}_{\ab \db}  {m}_{\bb \eb}  {m}_{\cb \fb} - {m}_{\ab \db}  {m}_{\bb \cb}  {m}_{\eb \fb}     \right) \ . 
\eeq
 That is to say we have reproduced exactly the potential for the scalar fields expected for gauged supergravity up to the term 
\beq
\Lambda^{\ab}{}_{\bb \cb \db} \Lambda^{\bb}{}_{\ab \eb \fb} \left( m^{\cb \eb } m^{\db \fb}- m^{\cb \db} m^{\eb \fb} \right)  \  ,
\eeq
which, after some algebra can be written as 
\beq
2 \pl_{kl} \left( m^{p k}  m^{\bar{q} \bar{l} } V_{\bar{q}}^{[q} \pl_{pq} V^{l] }_{\bar{l}} \right) \ , 
\eeq
and so is a total derivative and may be neglected.

It is worth remarking that the additional term in the Lagrangian $V_5$ was vital to achieve correct cancelations and contributions to this result.     
 
It is natural to ask whether the assumption that the trombone gauging vanishes is actually necessary; could one obtain an action principle for a trombone gauged supergravity?  From the above considerations it seem likely that an appropriate a scalar potential could be deduced. However, the trombone symmetry is only an on-shell symmetry of the full supergravity action and so to make such a conclusion it would be vital to include the other supergravity fields (i.e. the gauge and gravity sectors) in  a duality symmetric fashion.   

\subsection{Gauge invariance of the action} 

Under the local generalised diffeomorphism symmetry generated  by the derivative \eqref{sl5deriv} the action (before reduction) has the schematic variation 
\beq
\delta V =  G_0 + \dots 
\eeq 
in which the ellipsis indicates total derivative terms and $G_0$ is an expression which vanishes upon invoking the section condition.  To find the corresponding invariance constraint that the reduced theory must obey one could substitute the reduction ansatz into  $G_0 = 0$.   

However, a more economical and enlightening approach is to directly consider the variation of the reduced theory under the corresponding reduced symmetry \eqref{lred}.  
As we have seen, the reduction of the action produced (up to total derivatives) the potential for  gauged supergravity \eqref{Vgauged} which may be written in terms of $X_{mn,k}{}^l$ as follows
\begin{equation}
\begin{split}
V_{gauged}=&\fr{1}{64}\left(3X_{mn,r}{}^sX_{pq,s}{}^rm^{mp}m^{nq}-X_{mp,q}^{n}X_{nr,s}^{m}m^{pr}m^{qs}   \right)\\
&+\fr{1}{96}\left(X_{mn,r}{}^sX_{pq,t}{}^um^{mp}m^{nq}m^{rt}m_{su} + X_{mp,q}{}^nX_{nr,s}{}^mm^{pq}m^{rs}    \right).
\end{split}
\end{equation}
We have already stipulated a fundamental constraint  that these $X_{mn,k}{}^l$ are constant and, as we have seen, they may be understood as the generators of the gauge symmetry.  Thus the  $X_{mn,k}{}^l$ do not transform under gauge variation whilst the transformation of the metric $m^{ab}$ reads
\begin{equation}
\d_\xi m^{ab} = (X_{kl,m}{}^am^{mb}+X_{kl,m}{}^bm^{am})\xi^{kl}.
\end{equation}
It is clear then that the constraints of gauge invariance are exactly the standard ones obtained in gauged supergravity.  Explicitly, the gauge transformation of the action is
\begin{equation}
\delta_{\xi}V=\fr{1}{24}X_{a[b,c]}{}^dX_{dm,n}{}^{a}X_{kl,p}{}^bm^{cp}m^{mn}\xi^{kl}, 
\end{equation}
which becomes
\begin{equation}
\begin{split}
\delta_{\xi}V=&-\fr{1}{12}\e_{abcpq}Z^{pq,d}\d^{a}_{[d}Y_{m]n}X_{kl,p}{}^bm^{cp}m^{mn}\xi^{kl}\\
=&-\fr{1}{24}\left(\e_{abcpq}Z^{pq,a}Y_{mn}m^{mn} - \e_{abcpq}Z^{pq,d}Y_{dn}m^{an}\right)X_{kl,p}{}^bm^{cp}\xi^{kl}=0.
\end{split}
\end{equation}
The first term here is zero due to $Z^{[ab,c]}=0$ and the second is zero because of the quadratic constraint
\begin{equation}
Z^{mn,p}Y_{pq}=0.
\end{equation}

\subsection{SL(5) Summary}  
It is opportune at this stage to summarise. We have now seen that for the $SL(5)$ case the embedding tensor of $D=7$ maximal supergravities drops out upon Scherk--Schwarz reduction of the generalised Lie derivative.  Furthermore reduction of the action produces the known scalar potential of these supergravities.  Consistency of the reduction invokes a number of constraints:
\begin{enumerate}
\item The gaugings are constant and invariant.
\item The gauge transformations close onto the C-bracket.
\item The Jacobiator of gauge transformations generate trivial gauge transformations (which is automatic given 1 \& 2).
\item The action is invariant under the gauge transformations (which is automatic given 1 \& 2).
\end{enumerate}
It is evident that the strong constraint is a sufficient condition for these to hold however it not actually necessary. This is just as for the double field theory case \cite{Grana:2012rr}. A usual Kaluza--Klein reduction on the extended dimensions is equivalent to solving the strong constraint. A Scherk--Schwarz reduction is by definition one that has field dependence  on the new coordinates, albeit of a special type, and so can break the strong constraint. The fact that this produces a consistent theory i.e. gauged supergravity, indicates that one should allow dependence on these coordinates and move beyond just allowing trivial solutions to the physical section condition. This can happen however only in the very restricted circumstances that are described above.  

These constraints when applied to the normal Scherk--Schwarz reductions (where closure is not on the C--bracket but the usual Lie bracket etc.) essentially mean the internal manifold is locally isomorphic to a Lie group manifold.   What emerges here is the generalised geometric version of a group manifold with Lie derivatives being replaced by generalised Lie derivates and the algebra being the C--bracket rather than the Lie bracket.   Describing the properties of these manifolds will be an interesting direction for future study -- in  \cite{Hohm:2012gk}  recent progress  in this direction has  been made in understanding how to  ``exponentiate"  the local symmetries for the $O(d,d)$ case relevant to DFT.   A closely related direction for future study,  already carried out in the case of DFT in \cite{Dibitetto:2012rk},  would be to find explicit solutions of these constraints, in particular those for which the section condition {\it{is}} violated, since these would give a very clear geometric origin to supergravities that have no higher dimensional origin within the context of regular supergravity. 
 
\section{Generalisation to higher U--duality groups}
\subsection{$E_{d(d)}$ generalised geometry for $d=5,6,7$} 
As the dimensionality is increased,  one encounters higher dimensional U--duality groups from the exceptional series.  For all cases up to $E_{7(7)}$ the relevant  doubled geometry has been developed in \cite{Hull:2007zu,Pacheco:2008ps,Hillmann:2009ci,Berman:2010is,Berman:2011pe,Berman:2011jh,Berman:2011cg,Coimbra:2011ky} and is summarised in  table 1 using the notation of \cite{Coimbra:2011ky}.
\begin{table}[t]
  \centering
  \begin{tabular}{@{} c|ccc @{}}
    \hline
     & $\partial$ & N & Geralised Tangent Bundle \\ 
    \hline
    $SL(5)$ &  $\bf{10}$ & $\bf{\overline{5}}$   & $TM\oplus \Lambda^2 T^\ast M$ \\ 
    $SO(5,5)$ & \bf{16}  &  $\bf{10}$  &  $TM\oplus \Lambda^2 T^\ast M \oplus \Lambda^5T^{\ast}M $ \\ 
    $E_{6,{6}}$ & $\bf{27}$ & $\overline{\bf{27}}$   &     $TM\oplus \Lambda^2 T^\ast M \oplus \Lambda^5T^{\ast}M $  \\ 
     $E_{7,{7}}$ & $\bf{56}$ &  \bf{133}  &  $TM\oplus \Lambda^2 T^\ast M \oplus \Lambda^5T^{\ast}M \oplus(T^\ast M \otimes \Lambda^7 T^{\ast}M   )  $ \\ 
    \hline
  \end{tabular}
  \caption{ $\partial$ denotes the representation of the coordinates and $N$ that of the section condition $\partial \otimes_N \partial = 0 $. }
  \label{tab:label}
\end{table}

For all these cases (and indeed for the $O(d,d)$ DFT relevant to strings),  the appropriate derivative has the general form
\beq
\label{GenLie}
{\cal L}_\xi V^M = L_\xi V^M + Y^{MN}{}_{PQ} \pl_N \xi^P V^Q
\eeq
where $L_\xi V^M $ is the standard Lie derivative and $Y^{MN}{}_{PQ}$ is a group invariant tensor that essentially projects onto the section condition.  
For the U-duality groups up to $E_{7(7)}$ the structure of the derivatives was established in \cite{Coimbra:2011ky} and further investigated in  \cite{Martin} with the following results:\footnote{$\eta_{MN}$ is the invariant metric on $O(d,d)$; $\e_{i M N} = \e_{i mn , pq}$ is the $SL(5)$ alternating tensor;   $(\g^i)^{MN}$ are  $16\times16$ MW representation of the $SO(5,5)$ Clifford algebra (they are symmetric and the inverse $(\g^i)_{MN}$ is the inverse of $(\g^i)^{MN}$); $d^{MNR}$ is a symmetric invariant tensor of $E_6$ normalized such that $d^{MNP}\bar{d}_{MNP} = 27$; $c^{MNPQ}$ is a symmetric tensor of $E_7$  and $\epsilon^{MN}$ is the symplectic invariant  tensor of its {\bf 56} representation.  In all cases except $E_{7(7)}$, the tensor $Y$ is symmetric in both upper and lower indices. 
}
\bea
O(n,n)_{strings}: & \quad & Y^{MN}{}_{PQ}  = \eta^{MN} \eta_{PQ} \ , \non \\
SL(5):  & \quad & Y^{MN}{}_{PQ}= \e^{i MN}\e_{i PQ} \ , \non    \\
SO(5,5): &\quad & Y^{MN}{}_{PQ}  = \frac{1}{2} (\g^i)^{MN} (\g_i)_{PQ} \ ,  \non  \\
E_{6(6)}: &\quad & Y^{MN}{}_{PQ}  = 10 d^{MN R} \bar{d}_{PQR} \ ,\non   \\ 
E_{7(7)}: &\quad & Y^{MN}{}_{PQ}  = 12 c^{MN}{}_{PQ} + \delta^{(M}_P \delta^{N)}_Q + \frac{1}{2} \e^{MN} \e_{PQ } \ . \non  \\ 
\eea
These derivatives have a Jacobiator proportional to terms that vanish on the section condition.   One may solve the section condition by setting derivative corresponding to brane wrapping coordinates to zero.  On doing so the derivatives reduce to a derivative on sections of the generalised tangent bundle which can be thought of the $E_{n(n)}$ generalisation of the Dorfmann derivative on $TM \oplus T^{\ast} M$.  

As with the $Sl(5)$ case, the dynamical fields of the theory are packaged into a {\it generalised metric}, i.e.,  $G/H$ coset representative where $G$ is the appropriate U--duality group and $H$ is its maximal compact subgroup.  A systematic way of obtaining the appropriate generalised metric is presented in \cite{Berman:2011jh} using a non-linear realisation of the semi-direct product of $E_{11}$ and its first fundamental representation.\footnote{   A technicality: in this paper we consider generalised metrics scaled such that the top left hand entry (i.e. the part that acts on $TM \otimes TM$) contains the regular space time metric with no additional determinant factors.  For such a generalised metric the appropriate Lie derivative appearing  in \eqref{GenLie} is evidently that of a tensor of density weight zero.} 

In the following we shall restrict our attention to the case of $SO(5,5)$ and $E_{6(6)}$; the case of $E_{7(7)}$ is a little more complex -- as can be seen in the structure of the $Y$ tensor -- and we leave this for future work.

 \subsection{ Sherk-Schwarz and closure constraints}

  We now introduce a gauging by specifying a Scherk--Schwarz ansatz, $T^M_N(x, y) = W^M{}_{\bar{A}}(y)W^{\bar{B}}{}_{N}(y) T^{\bar{A}}_{\bar{B}}(x) $,  then we find that
\beq
{\cal L}_X V^M = W^M{}_{\bar{C}} \left[  {\cal L}_{\bar{X}} V^{\bar{C}} + F^{\bar{C}}{}_{\br{A}{B}}  V^{\bar{B}} X^{\bar{A}} \right]    \ , 
\eeq
with the {\it structure constants} given by the expression 
\beq
  F^{\bar{C}}{}_{\br{A}{B}} = W^{\bar{C}}{}_M \left[  \pl_{\bar{A}}W^M{}_{\bar{B}} -  \pl_{\bar{B}}W^M{}_{\bar{A}} + 
  Y^{\br{R}{D}}{}_{\br{S}{B}} W^M{}_{\bar{R}} \pl_{\bar{D}} W^N{}_{\bar{A}} W^{\bar{S}}{}_N  \right] \ . 
\eeq

The first thing to note is that, unlike the doubled field theory case, these structure constants are not anti-symmetric.  For cases upto $E_6$ we may express $Y^{MN}{}_{PQ} = \kappa  d^{a MN} d_{a PQ}$ where $d_{a PQ} = d_{a QP}$ is an invariant tensor  and $a$ is an index in the fundamental representation of the duality group.  Then if we define $V^a_m$ as the twist matrix in the vector representation we have the following
\bea
T^{\bar{C}}_{\br{A}{B}} &=& d^{\bar{a} \br{C}{D} } d_{\bar{a} \br{R} {A}} \pl_{\bar{D}} W^N{}_{\bar{B}}W^{\bar{R}}{}_N = 
d^{\bar{a} \br{C}{D}} d_{m NM} W^M{}_{\bar{A}} \pl_{\bar{D}} W^N{}_{\bar{B}} V_{\bar{a}}^m  \nonumber \\
&=& d^{\bar{a} \br{C}{D}} \left(  \pl_{\bar{D}}(d_{\bar{a} \br{B}{A}} ) - \pl_{\bar{D}}( d_{m N M} )   W^M{}_{\bar{A}}  W^N{}_{\bar{B}} V_{\bar{a}}^m  \right. \nonumber \\
&& \left. - d_{\bar{m} \br{B}{ A} }V^{\bar{m}}_m \pl_{\bar{D}} V_{\bar{a}}^m - 
d_{m N M}W^N{}_{\bar{B}} V_{\bar{a}}^m   \pl_{\bar{D}} W^M_{\bar{A}}   \right)  \nonumber \\
&=& - T^{\bar{C}}_{\br{B}{A}} -  d^{\bar{a} \br{C}{D}} d_{\bar{b} \br{B}{A} }V^{\bar{b}}_m \pl_D V^m_{\bar{a}} \ . 
\eea 
Then the symmetric piece to the gauging is given by:
\beq
F^{\bar{C}}{}_{\br{A}{B}} + F^{\bar{C}}{}_{\br{B}{A}} = - \kappa d^{\bar{a} \br{C}{D}} d_{\bar{b} \br{B}{A} }V^{\bar{b}}_m \pl_D V^m_{\bar{a}} \  .
\eeq

We have now some immediate constraints on the admissible gauging.  Firstly, these structure constants need to indeed be constant. Additionally we obtain constraints from the closure on the Courant bracket and from the Jacobiator.   Because the generalised derivative \eqref{GenLie} has a universal structure these have the same form as the constraints found for the $SL(5)$ case.  This is as expected and reflects the discussion in section 4.4 of  \cite{Coimbra:2011ky}. Essentially the key point is that the space admits a globally defined frame on the generalised tangent space and is thus ``generalised parallelizable". Given a global frame there is a globally defined basis with an identity structure. The embedding tensor then can be identified with the ``generalised torsion" which is the algebra (under the Dorfman derivative) of elements of this global basis.

\subsection{Twisted derivatives and the embedding tensor}
We now show that the above gaugings can be related to the embedding tensor construction of gauged supergravities.   This provides an explicit relation between the gaugings and the Scherk-Schwarz twisting.  In fact, the situation is in some regards easier than for the case of $SL(5)$ since the symmetric part of the gauging is sufficient to identify all components of the embedding tensor.   To avoid a mess of notation all indices with no accents in the following are flattened (we will drop the over barred notation) and curved indices, which rarely appear, are signaled by a hat.  
\subsubsection{$SO(5,5)$ gaugings} 
The embedding tensor of \cite{Bergshoeff:2007ef} is $\Theta_{M}^{\a} = \Theta_{M}^{[ij]} $ where $M$ is an index of the $\bf{16}$ and $\a = [ij]$ is an adjoint $\bf{45}$ index.  We have that 
\beq
\bf{ 16_{s} \otimes 45 = 16_{s} + 144_{c} + 560_{s}} \ . 
\eeq
Supersymmetry imposes a constraint on the embedding tensor $\mathbb{P}_{560} \Theta = 0$.  Components of the embedding tensor in the $\bf{16}$ are derived from the trombone gaugings.   The embedding tensor may then be parametrised by a gamma-traceless vector-spinor $Z^{Mi }$ ( $Z^{Mi } \gamma_{i MN} = 0$) and a spinor $\theta_{M}$.   In the adjoint  representation (i.e. adjoint of the gauge group) we have  
\beq
X_{MN}{}^{K} = - Z^{L i} \gamma^{j}_{L[M} \g_{ij N]}{}^{K}  - \frac{8}{5} \theta_{[M}\delta_{N]}^{K} + \gamma_{iMN} \left( - Z^{Ki} - \frac{2}{5} \g^{i KP}\theta_{P} \right) \ . 
\eeq
The relation 
\beq
(t_{\a})_{M}{}^{K}(t_{\a})_{N}{}^{L} = -\frac{1}{32} (\g^{{ij}})_{M }^{K} (\g_{{ij}})_{N}^{L} = \frac{1}{16} \delta_{M}^{K}\delta_{N}^{L} + \frac{1}{4}  \delta_{N}^{K}\delta_{M}^{L} - \frac{1}{8} \gamma_{iMN}\gamma^{iKL}
\eeq
ensures that the generalised derivative may be recast as 
\bea
L_{U}V& = & U^{N}\pl_{N} V^{M} -  V^{N}\pl_{N} U^{M}  + \frac{1}{2} \g_{iAB}\g^{iMS} \pl_{S} U^{A}V^{B } \nonumber \\
&=& U^{N}\pl_{N} V^{M}  + \frac{1}{2}   V^{M}\pl_{N} U^{N} - 4 (t_{\a})_{A}{}^{S}(t^{\a})_{B}{}^{M}  \pl_{S} U^{A}V^{B } \ . 
\eea
Pluging in the Scherk--Schwarz twist into the above bracket gives a gauging whose symmetric part is 
\beq
F_{(MN)}{}^{K} = - \frac{1}{4} \g^{i KQ} \g_{j MN } g^{j}_{\hat{i}} \pl_{Q} g_{i}^{\hat{i}}
\eeq
from which we find 
\beq
\theta_{L} = \frac{1}{16}\g^{j KP}\g_{i KL} g^{i}_{\hat{i}} \pl_{P} g_{j}^{\hat{i}}
\eeq
and
\beq
Z^{Ki} = \frac{1}{4} \left[\delta_{R}^{K}\delta_{k}^{i} - \frac{1}{10} \gamma^{i KP} \gamma_{k PR}   \right]  \g^{j RQ}  g^{k}_{\hat{i}} \pl_{Q} g_{j}^{\hat{i}} \  .
\eeq
In the above the $g$ is the group element in the vector representation of $SO(5,5)$.

\subsubsection{ $E_{6,(6)}$ gaugings}
The embedding tensor of \cite{deWit:2004nw} is $\Theta_{M}^{\a}$ where $M$ is an index of the $\bf{27}$ and $\a$ is an adjoint $\bf{78}$ index.  We have that 
\beq
\bf{ 27 \otimes 78 = 27 + 351 + \overline{1728}} \ . 
\eeq
Supersymmetry imposes a constraint on the embedding tensor $\mathbb{P}_{1728} \Theta = 0$.  Components of the embedding tensor in the $\bf{27}$ are derived from the trombone gaugings.   The embedding tensor may then be parametrised by an antisymmetric $Z^{PQ}$ and a vector $\theta_{M}$.   In the adjoint  representation (i.e. adjoint of the gauge group) we have  
\beq
X_{MN}{}^{K} = 10 Z^{PQ} d^{KAB}d_{NPA}d_{MQB} - \frac{3}{2} \theta_{[M}\delta_{N]}^{K} + d_{QMN} \left( Z^{QK} - \frac{15}{2} d^{QKP}\theta_{P} \right) \ . 
\eeq
In the above $d_{MNP}$ is the totally antisymmetric invariant tensor normalised such that $d_{MNQ} d^{MNP}= \delta_{P}^{Q}$.  

The  relation 
\beq
(t_{\a})_{M}{}^{K}(t_{\a})_{N}{}^{L} = \frac{1}{18} \delta_{M}^{K}\delta_{N}^{L} + \frac{1}{6}  \delta_{N}^{K}\delta_{M}^{L} - \frac{5}{3} d_{MNP}d^{KLP}
\eeq
ensures that the generalised derivative may be recast as 
\bea
L_{U}V& = & U^{N}\pl_{N} V^{M} -  V^{N}\pl_{N} U^{M}  + 10 d_{PAB}d^{PMS} \pl_{S} U^{A}V^{B } \nonumber \\
&=& U^{N}\pl_{N} V^{M}  + \frac{1}{3}   V^{M}\pl_{N} U^{N} - 6 (t_{\a})_{A}{}^{S}(t^{\a})_{B}{}^{M}  \pl_{S} U^{A}V^{B } \ . 
\eea

By comparing the symmetric part of $X_{MN}{}^{K}$ to the expressions for gaugings found by performing a Scherk--Schwarz twist in the brackets one see that 
\beq
\theta_{M} = \frac{2}{3} d_{T M R} d^{T PQ  }  G_{\hat{S}}^{R} \pl_{P} G_{Q}^{\hat{S}}
\eeq
and 
\beq
Z^{PQ} = 5 d^{UV [P} G^{Q]}_{\hat{S}} \pl_{U} G_{V}^{\hat{S}} \ . 
\eeq

\section{Conclusions and Discussion}

It is very satisfying to see how the Scherk--Shwarz reductions of the manifestly U--duality invariant form of M--theory produce the gauged supergravity potentials and how one can identify the embedding tensor in terms of the twist fields. Many of these ideas were present before in the literature; it is a test of this M--theory formalism that these ideas are realised exactly using the U-duality manifest actions. It also makes the novel coordinates play a more important role since a simple Kaluza Klein reduction in those directions is not carried out. Thus, the fields do depend on the new extra coordinates of the extended space. However, the full space should be ``generalised parallelisable", so the dependence on the new coordinates should be of a particular form. These constraints match those on the embedding tensor in gauged supergravity.

\section{Acknowledgements} 
DCT is supported by an FWO-Vlaanderen postdoctoral fellowship and this work is supported in part by the Belgian Federal Science Policy Office
through the Interuniversity Attraction Pole IAP VI/11 and by FWO Vlaanderen through project G011410N.  We would like to thank the Isaac Newton Institute, Cambridge and the participants of the BSM programme for inspiration whilst some of this work was completed. DSB is supported by an STFC rolling grant ST/J000469/1, String theory, gauge theory \& duality. ETM is supported by a Queen Mary fellowship. ETM acknowledges the warm hospitality at EMFCSC, Erice, Italy whilst some of this work was completed. DSB would like to especially thank Malcolm Perry for discussion on all things M-theory and to Dan Waldram for direct discussions on generalised geometry and twisting as well as Martin Cederwall, Olaf Hohm, Axel Kleinshmidt, the Godazgar bothers, Fabio Riccioni,  Henning Sambtleben and Peter West for general discussions relating to this work.

\begin{appendix}
\section{$SL(5)$ and $d=7$ gaugings} 
The generators of $SL(5)$, in the fundamental $\bf{5}$ representation, can be expressed as 
\beq
(T_a^b)^i_j  = \delta_a^i \delta_j^b - \frac{1}{5} \delta_a^b \delta^i_j  \   . 
\eeq
They are traceless, $Tr(T_a^b)= 0 $ and obey one relation $T_a^a = 0 $.  They obey the commutator relation
\beq
[T_a^b, T_c^d ] = \delta_c^b T_a^d-  \delta_a^d T_c^b \  . 
\eeq  
Also we have 
\beq
Tr_{\bf{5}}( T_a^b T_c^d) = - \delta_a^d \delta_c^b + \frac{1}{5} \delta_a^b \delta_c^d \ . 
\eeq
In the  $\bf{10}$ representation the generators are given by 
\beq
(T_a^b)^{ij}_{kl} = 2 (T_a^b)^{[i}_{[k} \delta_{l]}^{j]}    \ , 
\eeq
and obey the commutator
\beq
[T_a^b, T_c^d ] = (T_a^b)^{ij}_{kl} (T_c^d)^{kl}_{mn}  - (T_c^d)^{ij}_{kl} (T_a^b)^{kl}_{mn}    =  \delta_c^b T_a^d-  \delta_a^d T_c^b \  . 
\eeq  
We note the following useful identity
\beq
 (T_a^b)^{ij}_{kl} (T_b^a)^{rs}_{pq}  = - \frac{1}{4} \epsilon_{a klpq} \epsilon^{a ij rs} +\frac{1}{5} \delta^i_{[k} \delta^j_{l]}  \delta^r_{[p} \delta^s_{q]} + \delta^i_{[p} \delta^j_{q]}  \delta^r_{[k} \delta^s_{l]} \ . 
 \eeq
 This identity is used to establish that the following expressions for the generalised derivative are equivalent
 \def\pl{\partial}
 \def\non{\nonumber}
 \bea
{\cal L}_U V &=& U^N \pl_N V^M - V^N\pl_N U^M + \epsilon^{a MN }\epsilon_{a PQ} \pl_N U^P V^Q \non \\ 
&\equiv & \frac{1}{2}U^{ij}\pl_{ij} V^{kl} -   \frac{1}{2}V^{ij}\pl_{ij} U^{kl}  + \frac{1}{8}  \epsilon^{a ij kl }\epsilon_{a pq rs} \pl_{ij} U^{pq} V^{rs}    \\ 
{\cal L}_U V  &=& \frac{1}{2}U^{ij}\pl_{ij} V^{kl}  - \frac{1}{2} (T_a^b)^{ij}_{pq} (T^a_b)^{kl}_{rs} \pl_{ij} U^{pq} V^{rs} + \frac{1}{10} V^{kl}  \pl_{ij} U^{ij}  \label{WaldramForm} \\ 
{\cal L}_U V  &=&  \frac{1}{2}U^{ij}\pl_{ij} V^{kl}   + \frac{1}{2} V^{kl} \pl_{ij} U^{ij} + V^{ki}\pl_{ij}U^{j l} - V^{li}\pl_{ij}U^{kl} \ . 
 \eea
 The first form shows how the derivative fits in to the general group structure applicable across dimensions, the second makes explicit the relation to the derivatives that appear in \cite{Coimbra:2011ky} and the final one is the form given in \cite{Berman:2011cg}.

The embedding tensor for $d=7$ gauged supergravity, with no trombone gauging, is given by  \cite{Samtleben:2005bp}:
\beq
\Theta_{mn, p}{}^q = \delta^q_{[m} Y_{n]p} - 2 \epsilon_{mn p rs}Z^{rs ,q} 
\eeq
with $Y_{mn}= Y_{(mn)}$ in the $\bf{15}$ and $Z^{rs, q} = Z^{[rs],q}$ in the $\bf{\overline{40}}$ so that $Z^{[rs, q] }  = 0$.  
It is traceless  $\Theta_{mn, p}{}^p = 0 $ and hence the gauge group generators in the $\bf{5}$ and $\bf{10}$ are given by
\beq
X_{mn , p }{}^q = \Theta_{mn, p}{}^q \ , \quad 
X_{mn,  pq }{}^{rs} = 2 \Theta_{mn, [p}{}^{[r}\delta^{s]}_{q]} \ .  
\eeq

To incorporate the trombone gauging we  introduce an extra generator $(T_0)_p^q = \delta_p^q $ corresponding to the $\mathbb{R}^{+}$ and propose an ansatz (we follow exactly  \cite{LeDiffon:2008sh} where the procedure is carried out for all the other exceptional groups)
\bea
\hat{\Theta}_{mn, 0} &=& \theta_{mn} \ , \non  \\
\hat{\Theta}_{mn, p}{}^q &=& \delta^q_{[m} Y_{n]p} - 2 \epsilon_{mn p rs}Z^{rs ,q}  + \zeta \theta_{ij} (T_p^q)^{ij}_{mn}  \ ,
\eea
where $\theta_{mn} = \theta_{[mn]}$ is in the $\bf{10}$. 
Then the gauge generators in the fundamental are given by 
\bea
\hat{X}_{mn, p}{}^q &=&  \hat{\Theta}_{mn, 0} (T_0)^q_p + \hat{\Theta}_{mn , r}{}^s (T_r^s)_p^q   \non  \\ 
&=& \delta^q_{[m} (Y_{n]p} -2 \zeta \theta_{n] p}  ) - 2 \e_{mn p rs}Z^{rs ,q}  +\frac{1}{5} (5-2 \zeta) \theta_{mn} \delta_p^q,
\eea
and in the ${\bf 10}$ by 
\bea
\hat{X}_{mn, pq}{}^{rs} &=&  \hat{\Theta}_{mn, 0} (T_0)^{rs}_{pq} + \hat{\Theta}_{mn , a }{}^b (T_b^a)^{rs}_{pq}   \non \\ 
&=& 2 \Theta_{mn, [p}{}^{[r}\delta^{s]}_{q]} +  2\theta_{mn} \delta^{[r}_{[p} \delta^{s]}_{q]}   + \zeta  \theta_{ij} (T_a^b)^{ij}_{mn} (T^b_a)^{rs}_{pq} \non \\
&=& 2 \Theta_{mn, [p}{}^{[r}\delta^{s]}_{q]} +  \left(2 + \frac{\zeta}{5}  \right)\theta_{mn} \delta^{[r}_{[p} \delta^{s]}_{q]}  + \zeta \theta_{pq} \delta^{[r}_{[m} \delta^{s]}_{n]}   -\frac{1}{4} \zeta  \theta_{ij}\epsilon^{ij rs a} \epsilon_{a mn pq}  \non \\
\eea
One now calculates the symmetric part of the gauging 
\beq
 \hat{X}_{mn, pq}{}^{rs}  + \hat{X}_{pq, mn}{}^{rs}    =2 \epsilon_{a mn pq}\left( Z^{rs ,a}  - \frac{\zeta}{4} \epsilon^{rs a i j }\theta_{ij} \right)  +    \left( 2 + \frac{6 \zeta}{5} \right) \left(\theta_{mn} \delta^{[r}_{[p} \delta^{s]}_{q]} +\theta_{pq} \delta^{[r}_{[m} \delta^{s]}_{n]}  \right) \  .
\eeq
The requirements of supersymmetry as explained in \cite{LeDiffon:2008sh} are that this falls in  the same representation as without the trombone gauging hence we fix   $\zeta = - \frac{5}{3}$.    Although the symmetric part of the gauging does not depend on $Y$ note that the  that the antisymmetric part of the gauging  depends on $\theta$ $Z$ and $Y$.

\end{appendix}
 
\providecommand{\href}[2]{#2}\begingroup\raggedright\endgroup

\end{document}